\documentclass[fleqn]{2023SCGE}
\usepackage{orcidlink}

\usepackage{multirow}
\newcommand{\nc}{\newcommand*}
\nc{\mU}{{\mathcal{U}}}
\nc{\yr}{\mathrm{yr}}

\nc{\red}[1]{\textcolor{red}{#1}}
\nc{\Eq}[1]{Eq.~\eqref{#1}}     
\nc{\Fig}[1]{Fig.~\ref{#1}}     
\nc{\Table}[1]{Table~\ref{#1}}  
\nc{\Sec}[1]{Sec.~\ref{#1}}     

\begin{document}

\ensubject{subject}

\ArticleType{Article}
\SpecialTopic{SPECIAL TOPIC: }
\Year{2024}
\Month{}
\Vol{}
\No{}
\DOI{}
\ArtNo{}
\ReceiveDate{}
\AcceptDate{}
\OnlineDate{}

\title{Can we distinguish between adiabatic and isocurvature fluctuations with pulsar timing arrays?}{Can we distinguish between adiabatic and isocurvature fluctuations with pulsar timing arrays?}

\author[1,2]{Zu-Cheng~Chen\orcidlink{0000-0001-7016-9934}}{}
\author[3]{Lang~Liu\orcidlink{0000-0002-0297-9633}}{liulang@bnu.edu.cn}

\AuthorMark{Z.-C. Chen}
\AuthorCitation{Z.-C. Chen, L. Liu}

\address[1]{Department of Physics and Synergetic Innovation Center for Quantum Effects and Applications, Hunan Normal University, Changsha, Hunan 410081, China}
\address[2]{Institute of Interdisciplinary Studies, Hunan Normal University, Changsha, Hunan 410081, China}
\address[3]{Faculty of Arts and Sciences, Beijing Normal University, Zhuhai 519087, China}


\abstract{
Understanding the nature of primordial fluctuations is pivotal to unraveling the Universe’s early evolution. While these fluctuations are observed to be nearly scale-invariant, quasi-adiabatic, and Gaussian on large scales, their small-scale behavior remains poorly constrained, offering a potential window into new physics. Recent detections of a stochastic gravitational wave background in the nanohertz frequency range by pulsar timing arrays (PTAs), including NANOGrav, PPTA, EPTA+InPTA, and CPTA, align with astrophysical predictions from supermassive black hole binaries but could also encode signatures of primordial phenomena. We investigate whether the observed signal originates from primordial isocurvature or adiabatic fluctuations by fitting them to the latest NANOGrav dataset. Through comprehensive Bayesian model comparison, we evaluate the distinguishability of these scenarios given current PTA sensitivities. Our results demonstrate that existing data cannot conclusively differentiate between isocurvature and adiabatic sources, highlighting the need for enhanced observational capabilities to probe the primordial universe at small scales.
}

\maketitle

\begin{multicols}{2}
\section{Introduction}\label{section1}

Adiabatic fluctuations and isocurvature fluctuations represent two distinct types of primordial density perturbations that can arise in the early Universe. Adiabatic fluctuations manifest as perturbations in the overall density of matter and energy throughout the Universe~\cite{Akrami:2018odb}. These perturbations uniformly impact all forms of matter and radiation, leading to a consistent shift in the density distribution. 
In simpler terms, the relative proportions of various constituents (e.g., dark matter, baryons, and radiation)\Authorfootnote
remain constant within the perturbed regions. 
Isocurvature fluctuations, on the other hand, denote perturbations in the relative abundances of different matter and energy components within the Universe~\cite{Kodama:1985bj,Bucher:1999re}. Unlike adiabatic fluctuations, they do not change the overall density. Instead, they modify the relative ratios between distinct types of matter or radiation, leading to a change in the composition of the Universe without affecting its total density.

Our Universe is characterized by primordial fluctuations that exhibit certain properties on large scales. These fluctuations are remarkably small, nearly scale-invariant, quasi-adiabatic, and almost Gaussian. Their presence is directly observed through the anisotropy pattern of the cosmic microwave background (CMB)~\cite{COBE:1992syq,Planck:2018jri}. However, our knowledge and constraints regarding the state of the Universe on small scales, where nonlinear physics processes primordial information, are much more limited. When examining large scales, the CMB provides valuable insights, indicating that isocurvature primordial fluctuations likely contribute no more than $1-10\%$ of the total fluctuations~\cite{Planck:2018jri}. Considering that the measured amplitude of the power spectrum of primordial adiabatic fluctuations is approximately $10^{-9}$, it follows that the power spectrum of isocurvature fluctuations may have an amplitude of less than $10^{-10}$ on large scales. However, the situation changes when we shift our focus to small scales, where the constraints imposed by the CMB no longer hold. On these smaller scales, our most promising avenue for constraining the amplitude and nature of primordial fluctuations lies in the study of primordial black holes (PBHs)~\cite{Zeldovich:1967lct,Hawking:1971ei,Carr:1974nx} and induced gravitational waves (IGWs)~\cite{Ananda:2006af,Baumann:2007zm,Garcia-Bellido:2016dkw,Inomata:2016rbd,Garcia-Bellido:2017aan,Kohri:2018awv,Cai:2018dig}. To investigate the early Universe's nature further, pulsar timing arrays (PTAs) prove to be indispensable tools. PTAs possess sensitivity in the nHz frequency band, enabling them to probe the characteristics of the early Universe in a unique and valuable manner.

Recently, several PTA collaborations, including the North American Nanohertz Observatory for Gravitational Waves (NANOGrav)~\cite{NANOGrav:2023gor,NANOGrav:2023hde}, the Parkers PTA (PPTA)~\cite{Zic:2023gta,Reardon:2023gzh}, the European PTA (EPTA) in conjunction with the Indian PTA (InPTA)~\cite{EPTA:2023sfo,Antoniadis:2023ott}, and the Chinese PTA (CPTA)~\cite{Xu:2023wog}, have collectively unveiled compelling evidence supporting the existence of a stochastic gravitational wave background (SGWB) within the frequency range of approximately $1 - 10\,\mathrm{nHz}$, which attracts lots of implications~\cite{NANOGrav:2023hvm,Antoniadis:2023xlr,King:2023cgv,Niu:2023bsr,Bi:2023tib, Vagnozzi:2023lwo,Fu:2023aab,Han:2023olf,Li:2023yaj,Shen:2023pan,Murai:2023gkv,Li:2023bxy,Anchordoqui:2023tln,Ghosh:2023aum,Wu:2023hsa,Bian:2023dnv,Li:2023tdx,Geller:2023shn,Ye:2023xyr,ValbusaDallArmi:2023nqn,DeLuca:2023tun,Choudhury:2023kam,Gorji:2023sil,Das:2023nmm,Ellis:2023oxs,He:2023ado,Kawasaki:2023rfx,King:2023ayw,Maji:2023fhv,Zhu:2023lbf,Huang:2023chx,Jiang:2023gfe,DiBari:2023upq,Aghaie:2023lan,Garcia-Saenz:2023zue,InternationalPulsarTimingArray:2023mzf,Altavista:2023zhw,Lozanov:2023rcd,Cang:2023ysz}. While the inferred amplitude and spectrum of the SGWB align with astrophysical expectations for a signal originating from the population of supermassive black hole binaries (SMBHBs), the possibility of new physics within this observational window remains a tantalizing prospect. One plausible explanation for the observed SGWB is that it arises from IGWs stemming from primordial fluctuations~\cite{Dandoy:2023jot,Franciolini:2023pbf,Franciolini:2023wjm,Inomata:2023zup,Cai:2023dls,Wang:2023ost,Liu:2023ymk,Unal:2023srk,Figueroa:2023zhu,Yi:2023mbm,Zhu:2023faa,Firouzjahi:2023lzg,Li:2023qua,You:2023rmn,Balaji:2023ehk,HosseiniMansoori:2023mqh,Zhao:2023joc,Liu:2023pau,Yi:2023tdk,Bhaumik:2023wmw,Choudhury:2023hfm,Yi:2023npi,Harigaya:2023pmw,Jin:2023wri,Cannizzaro:2023mgc,Zhang:2023nrs,Liu:2023hpw,Choudhury:2023fwk,Tagliazucchi:2023dai,Basilakos:2023jvp,Inomata:2023drn,Li:2023xtl,Domenech:2023dxx,Gangopadhyay:2023qjr,Cyr:2023pgw,Chen:2024fir,Choudhury:2023fjs,Choudhury:2024one}. Furthermore, cosmological phenomena such as first-order phase transitions~\cite{Addazi:2023jvg,Athron:2023mer,Zu:2023olm,Jiang:2023qbm,Xiao:2023dbb,Abe:2023yrw,Gouttenoire:2023bqy,An:2023jxf,Chen:2023bms} and the presence of topological defects like cosmic strings~\cite{Chen:2022azo,Kitajima:2023vre,Ellis:2023tsl,Wang:2023len,Antusch:2023zjk,Ahmed:2023pjl,Basilakos:2023xof,Chen:2023zkb} and domain walls~\cite{Kitajima:2023cek,Blasi:2023sej,Babichev:2023pbf} could also serve as sources for generating the SGWB. The SGWB generated by these cosmological sources holds immense potential for advancing our understanding of physics beyond the Standard Model and providing invaluable insights into the early stages of the Universe.

In this study, our goal is to investigate the potential explanations for the observed signals through IGWs stemming from both primordial isocurvature and adiabatic fluctuations. To achieve this, we utilize parameterized primordial spectra for both types of fluctuations to effectively match the latest observations reported by the NANOGrav. Moreover, we calculate the Bayes factors to compare different models. Our analysis indicates that, with the current sensitivity of PTAs, it is challenging to differentiate between the contributions of isocurvature and adiabatic fluctuations. The structure of this paper is outlined as follows: In Section~\ref{SIGW}, we introduce the general formalisms for IGWs arising from both types of primordial fluctuations. {In Section~\ref{PBH}, we provide an overview of the abundance of PBH from both types of primordial fluctuations}.
Section~\ref{Data} details the methodology employed in our data analysis and presents the findings using the latest NANOGrav data set, along with the computed Bayes factors for comparing different models. We conclude by summarizing our results and discussing their implications in Section~\ref{Con}.

\section{IGWs from primordial isocurvature fluctuations and primordial adiabatic fluctuations} 
\label{SIGW}

For the SGWB to be detectable, it is crucial for the primordial perturbations to undergo significant amplification compared to the fluctuations observed in CMB experiments. The amplitude and shape of the SGWB spectrum are influenced by various factors associated with the primordial fluctuations. These factors encompass the power spectrum and non-Gaussianities linked to the fluctuations, as well as the initial conditions of the primordial perturbations. Together, these elements shape the characteristics of the SGWB spectrum, determining both its amplitude and shape. 
In this section, we will provide an overview of the general formalisms concerning IGWs arising from primordial isocurvature fluctuations and primordial adiabatic fluctuations. The formalism for analytically calculating GWs induced by primordial adiabatic fluctuations during the radiation-dominated epoch was initially established in Refs.~\cite{Matarrese:1992rp, Matarrese:1993zf, Ananda:2006af, Baumann:2007zm, Espinosa:2018eve, Kohri:2018awv} and reviewed in Ref.~\cite{Domenech:2021ztg}. Conversely, the formalism for analytically calculating GWs induced by primordial isocurvature fluctuations during the radiation-dominated epoch was introduced in Refs.~\cite{Domenech:2021and} and reviewed in Ref.~\cite{Domenech:2023jve}.

The characterization of SGWBs today often involves describing their energy density per logarithmic frequency interval relative to the critical density $\rho_{c}$,
\begin{equation}
\Omega_{\mathrm{GW}}(k) \equiv \frac{1}{\rho_c} \frac{\mathrm{d} \rho_{\mathrm{GW}}(k)}{\mathrm{d} \ln k}.
\end{equation}
After generation, the energy density of GWs evolves in the same way as radiation. Utilizing this property, it is straightforward to determine the energy density of GWs at present. Taking the late-time limit during the radiation-dominated era, the relation of the GW energy density at present ($\Omega_{\rm GW}$) and the generation ($\Omega_{\rm GW,c}$) is given by \cite{Ando:2017veq}
	\begin{align}\label{eq:spectraldensitytoday2}
	\Omega_{\rm GW}(k)&= \Omega_{r,0} \left(\frac{g_{*,r}(T)}{g_{*,r}(T_0)}\right)\left(\frac{g_{*,s}(T)}{g_{*,s}(T_0)}\right)^{-4/3}\Omega_{\rm GW,c}(k)\,
\notag \\ &\approx 0.4 \left( \frac{g_{*,r}(T)}{80}\right) \left( \frac{g_{*,s}(T)}{80}\right)^{-4/3}  \Omega_{r,0}\, \Omega_{\rm GW,c}(k)\,
	\end{align}
where $\Omega_{r,0}$ is the density fraction of radiation in the present epoch, $g_{*,r}(T_c)$ and $g_{*s}(T_c)$ are the effective number of degrees of freedom in the energy density and entropy, respectively. The values today are denoted as $g_{*,r}(T_0) = 3.36$ and $g_{*,s}(T_0) = 3.91$. For the temperature dependence of $g_{*,r}(T)$ and $g_{*, s}(T)$, see Ref.~\cite{Saikawa:2018rcs}.

Following the Refs. \cite{Kohri:2018awv,Espinosa:2018eve}, the energy density of IGWs arising from primordial adiabatic fluctuations at the epoch of matter-radiation equality can be presented as
\begin{equation}
\Omega_{\mathrm{GW,c}}(k) = \int_0^{\infty} \mathrm{d} v \int_{|1-v|}^{1+v} \mathrm{d} u \mathcal{T}_{\zeta}(u, v) {\mathcal{P}}_{\zeta}(ku) {\mathcal{P}}_{\zeta}(kv),
\end{equation}
where ${\mathcal{P}}_{\zeta}$ is the primordial power spectrum of curvature perturbations and 
 \begin{equation}
\begin{aligned}
\mathcal{T}_{\zeta}(u,v)= & \frac{3}{1024 v^8 u^8}\left[4 v^2-\left(v^2-u^2+1\right)^2\right]^2\left(v^2+u^2-3\right)^2 \\
& \times\bigg\{\left[\left(v^2+u^2-3\right) \ln \left(\left|\frac{3-(v+u)^2}{3-(v-u)^2}\right|\right)-4 v u\right]^2 \\
& +\pi^2\left(v^2+u^2-3\right)^2 \Theta(v+u-\sqrt{3})\bigg\}.
\end{aligned}
\end{equation}

Under the assumption of Gaussian isocurvature fluctuations, we can derive a concise expression for the spectral density of IGWs originating from primordial isocurvature fluctuations \cite{Domenech:2021and}. This expression is given by
    \begin{align}\label{eq:Phgaussian}
    \Omega_{\rm GW,c}(k)=\int_0^\infty dv\int_{|1-v|}^{1+v}du  \mathcal{T}_s(\kappa, u, v)  {{\mathcal{P}}_{s}(ku)}{{\mathcal{P}}_{s}(kv)}\,.
    \end{align}
Here, we define $\kappa \equiv k/k_{\rm eq}$, where $k_{\rm eq}$ represents the comoving wavenumber at radiation-matter equality. ${\mathcal{P}}_{s}(k)$ corresponds to the  dimensionless spectrum of isocurvature fluctuations. The transfer function $\mathcal{T}_s$ is given by
\begin{equation}
\mathcal{T}_s(\kappa, u, v)=\frac{1}{3} \left(\frac{4v^2-(1-u^2+v^2)^2}{4uv}\right)^2 \left(I_{c,\infty}^2 + I_{s,\infty}^2\right)\,,
\end{equation}
where 
\begin{align}
I_{c,\infty}&(\kappa,u,v)=\frac{9}{32u^4v^4\kappa^{2}}\Bigg\{\left(-3+v^2\right)\left(-3+v^2+2u^2\right)\ln\left|1-\frac{v^2}{{3}}\right|\nonumber\\&
	-3u^2v^2+\left(-3+u^2\right)\left(-3+u^2+2v^2\right)\ln\left|1-\frac{u^2}{{3}}\right|\nonumber\\&
	-\frac{1}{2}\left(-3+v^2+u^2\right)^2\ln\left[\left|1-\frac{(u+v)^2}{{3}}\right|\left|1-\frac{(u-v)^2}{{3}}\right|\right]\Bigg\}\,,
\end{align}
and
\begin{align}
	I_{s,\infty}&(\kappa,u,v)=\frac{9\pi}{32u^4v^4 \kappa^{2}}\Bigg\{9-6v^2-6u^2+2u^2v^2\nonumber\\&+\left(3-u^2\right)\left(-3+u^2+2v^2\right)\Theta\left(1-\frac{u}{\sqrt{3}}\right)\nonumber\\&
	+\left(3-v^2\right)\left(-3+v^2+2u^2\right)\Theta\left(1-\frac{v}{\sqrt{3}}\right)\nonumber\\&
	+\frac{1}{2}\left(-3+v^2+u^2\right)^2\left[\Theta\left(1-\frac{u+v}{\sqrt{3}}\right)+\Theta\left(1+\frac{u-v}{\sqrt{3}}\right)\right]\Bigg\}\,.
\end{align}

\section{PBHs from primordial isocurvature fluctuations and primordial adiabatic fluctuations}
\label{PBH}
The production of IGWs depends on amplified scalar perturbations at small scales, which can concurrently result in the formation of PBHs. This section introduces the formalisms for PBHs originating from sizable primordial isocurvature and adiabatic fluctuations. The total fraction of PBHs in the present-day dark matter can be defined as
\begin{equation}
f_{\mathrm{PBH}} \equiv \frac{\Omega_{\mathrm{PBH}}}{\Omega_{\mathrm{DM}}}=\int \frac{d M}{M} f(M),
\end{equation}
where $\Omega_{\mathrm{DM}}$ represents the dark matter fraction and $f(M)$ denotes the PBH mass function. The horizon mass $M_H$ and the wavenumber $k$ are related by the following expression:
\begin{equation}
M_H \simeq \gamma M_{\odot}\left(\frac{g_{*,s}(T)}{10.75}\right)^{-2 / 3}\left(\frac{g_{*,r}(T)}{10.75}\right)^{1 / 2}\left(\frac{4.2 \times 10^6 \mathrm{Mpc}^{-1}}{k}\right)^2.
\end{equation}
\begin{table*}[t]
\centering
\caption{Parameters along with their respective prior distributions utilized in the Bayesian inference analysis, where $\mathcal{U}$ denotes a uniform distribution. 
Additional physical constraints are imposed requiring $f_{\rm PBH} \leqslant 1$ to ensure PBH abundance does not exceed dark matter density (see \Sec{PBH} for more details), as well as that the constraints from CMB \cite{Planck:2018vyg} and BBN \cite{Cooke:2013cba}.
Results are reported as the median and the $90\%$ credible interval, with symmetrical tails.}
\label{table:priors}
	\begin{tabular}{cclll}
		\hline
		\textbf{Model} & \textbf{Parameter} & \textbf{Description} & \textbf{Prior} & \textbf{Results} \\
        \hline
		\multirow{2}{*}{$M_0$} & \multicolumn{3}{c}{\textit{SMBHB}} \\[1pt]
		& $\log_{10} A_{\mathrm{SMBHB}}$ & Amplitude of the power spectrum. & $\mU(-15, -12)$ & $-14.58^{+0.09}_{-0.15}$\\
		\hline
		\multirow{3}{*}{$M_1$} & \multicolumn{3}{c}{\textit{Adiabatic Delta}} \\[1pt]
		& $\log_{10} A$ & Amplitude of the power spectrum. & $\mU(-3, 0)$ & $-1.46^{+0.21}_{-0.34}$\\
		& $\log_{10} (f_*/\mathrm{Hz})$ & Pivot frequency. & {$\mU(-8, -5)$} & $-6.74^{+0.32}_{-0.51}$\\
		\hline
		\multirow{4}{*}{$M_2$} &\multicolumn{3}{c}{\textit{Adiabatic Lognormal}} \\[1pt]
		& $\log_{10} A$ & Amplitude of the power spectrum. & {$\mU(-3, 0)$} & {$-0.81^{+0.48}_{-0.70}$}\\
		& $\log_{10} (f_*/\mathrm{Hz})$ & Pivot frequency. & {$\mU(-8, -2)$} & {$-6.71^{+1.11}_{-0.79}$}\\
        & $\Delta$ & Width of the power spectrum. & {$\mU(0.01, 5)$} & {$1.18^{+1.84}_{-1.02}$}\\
		\hline
		\multirow{4}{*}{$M_3$} &\multicolumn{3}{c}{\textit{Adiabatic Box}} \\[1pt]
		& $\log_{10} A$ & Amplitude of the power spectrum. & {$\mU(-3, 0)$} & {$-1.46^{+0.77}_{-0.47}$}\\
		& $\log_{10} (f_\mathrm{min}/\mathrm{Hz})$ & Minimum frequency of the power spectrum. & {$\mU(-9, -4)$} & {$-7.57^{+0.67}_{-0.59}$} \\
        & $\log_{10} (f_\mathrm{max}/\mathrm{Hz})$ & Maximum frequency of the power spectrum. & {$\mU(-8, -2)$} & {$-5.45^{+3.02}_{-1.60}$}\\
		\hline
		\multirow{3}{*}{$M_4$} & \multicolumn{3}{c}{\textit{Isocurvature Delta}} \\[1pt]
		& $\log_{10} A$ & Amplitude of the power spectrum. & {$\mU(16, 20)$} & {$17.9^{+0.7}_{-1.2}$}\\
		& $\log_{10} (f_*/\mathrm{Hz})$ & Pivot frequency. & {$\mU(-8, -6)$} & {$-6.88^{+0.30}_{-0.46}$}\\
		\hline
		\multirow{4}{*}{$M_5$} & \multicolumn{3}{c}{\textit{Isocurvature Lognormal}} \\[1pt]
		& $\log_{10} A$ & Amplitude of the power spectrum. & {$\mU(10, 30)$} & {$21.1^{+4.4}_{-3.9}$}\\
		& $\log_{10} (f_*/\mathrm{Hz})$ & Pivot frequency. & {$\mU(-8, -2)$} & {$-4.88^{+2.63}_{-2.37}$}\\
        & $\Delta$ & Width of the power spectrum. & {$\mU(0.01, 5)$} & {$1.39^{+0.64}_{-1.18}$}\\
		\hline
		\multirow{4}{*}{$M_6$} & \multicolumn{3}{c}{\textit{Isocurvature box}} \\[1pt]
		& $\log_{10} A$ & Amplitude of the power spectrum. & {$\mU(15, 25)$} & {$18.0^{+2.6}_{-1.1}$}\\
		& $\log_{10} (f_\mathrm{min}/\mathrm{Hz})$ & Pivot frequency. & {$\mU(-8, -6)$} & {$-7.18^{+0.83}_{-0.37}$}\\
        & $\log_{10} (f_\mathrm{max}/\mathrm{Hz})$ & Width of the power spectrum. & {$\mU(-7, -2)$} & {$-4.54^{+2.27}_{-2.13}$}\\
		\hline
	\end{tabular}	
\end{table*}
According to Refs.~\cite{Inomata:2017okj, Ando:2018qdb}, the mass function of PBHs arising from primordial adiabatic fluctuations can be expressed as
\begin{equation}
f(M) \simeq \gamma^{3 / 2}\left(\frac{\beta(M)}{1.6 \times 10^{-9}}\right)\left(\frac{10.75}{g_{*,r}(T)}\right)^{-3 / 4}\left(\frac{10.75}{g_{*,s}(T)}\right)\left(\frac{0.12}{\Omega_{\mathrm{DM}} h^2}\right)\left(\frac{M_{\odot}}{M}\right)^{1 / 2},
\end{equation}
where $\gamma \simeq 0.4$ is the ratio between the PBH mass $M$ and the horizon mass $M_{H}$~\cite{Niemeyer:1997mt}, \emph{i.e.}, $M=\gamma M_{H}$, and $\beta(M)$ is the the PBH production rate. Assuming a Gaussian distribution for the density perturbations, the PBH production rate can be calculated by integrating the Gaussian distribution of perturbations, namely
\begin{equation}
\beta(M)=\int_{\delta_c} \frac{d \delta}{\sqrt{2 \pi} \sigma_{\zeta}(M)} \exp \left(-\frac{\delta^2}{2 \sigma_{\zeta}^2(M)}\right),
\end{equation}
where $\delta_{c} \simeq 0.45$ is the critical density contrast equired for PBH formation~\cite{Musco:2004ak}, and $\sigma_{\zeta}^{2}(M)$ is the variance of the density contrast, which in the radiation domination era can be given by
\begin{equation}
\sigma_{\zeta}^2(M)=\frac{16}{81} \int \frac{d q}{q}\left(\frac{q}{k(M)}\right)^4 W^2\left(\frac{q}{k(M)}\right) T^2\left(q, k^{-1}(M)\right) \mathcal{P}_{\zeta}(q).
\end{equation}
Here $W(x)=e^{-x^2/2}$ is the Fourier transform of a volume-normalized Gaussian window smoothing function, and 
\begin{equation}
\mathcal{T}\left(q, k^{-1}\right)=3\left[\sin \left(\frac{q}{\sqrt{3} k}\right)-\left(\frac{q}{\sqrt{3} k}\right) \cos \left(\frac{q}{\sqrt{3} k}\right)\right] \left(\frac{q}{\sqrt{3} k}\right)^{-3}    
\end{equation}
denotes the transfer function.
Following Refs.~\cite{Passaglia:2021jla, Domenech:2023jve}, the mass function of PBHs originating from primordial isocurvature fluctuations can be expressed as
\begin{equation}
 f(M) \simeq \frac{  3^{q/2} k_\text{eq}^{2 q} \sigma_{s} ^{2q+1}(M)b}{ 2^q k^{2 q} \xi },  
\end{equation}
where $b\simeq 0.02$, $q \simeq 13/2$ , $\xi \simeq 1$ are constants which encode the PBH formation probability \cite{Passaglia:2021jla}, and the variance of the fluctuation $\sigma_s(M)$ is given by
\begin{equation}
    \sigma_s(M)\simeq\int d \ln(q) \left(\frac{q}{k(M)}\right)^4 W^2(\frac{q}{k(M)})\mathcal{P}_{s}(q).
\end{equation}

\begin{figure}[H]
\centering
\includegraphics[width=0.45\textwidth]{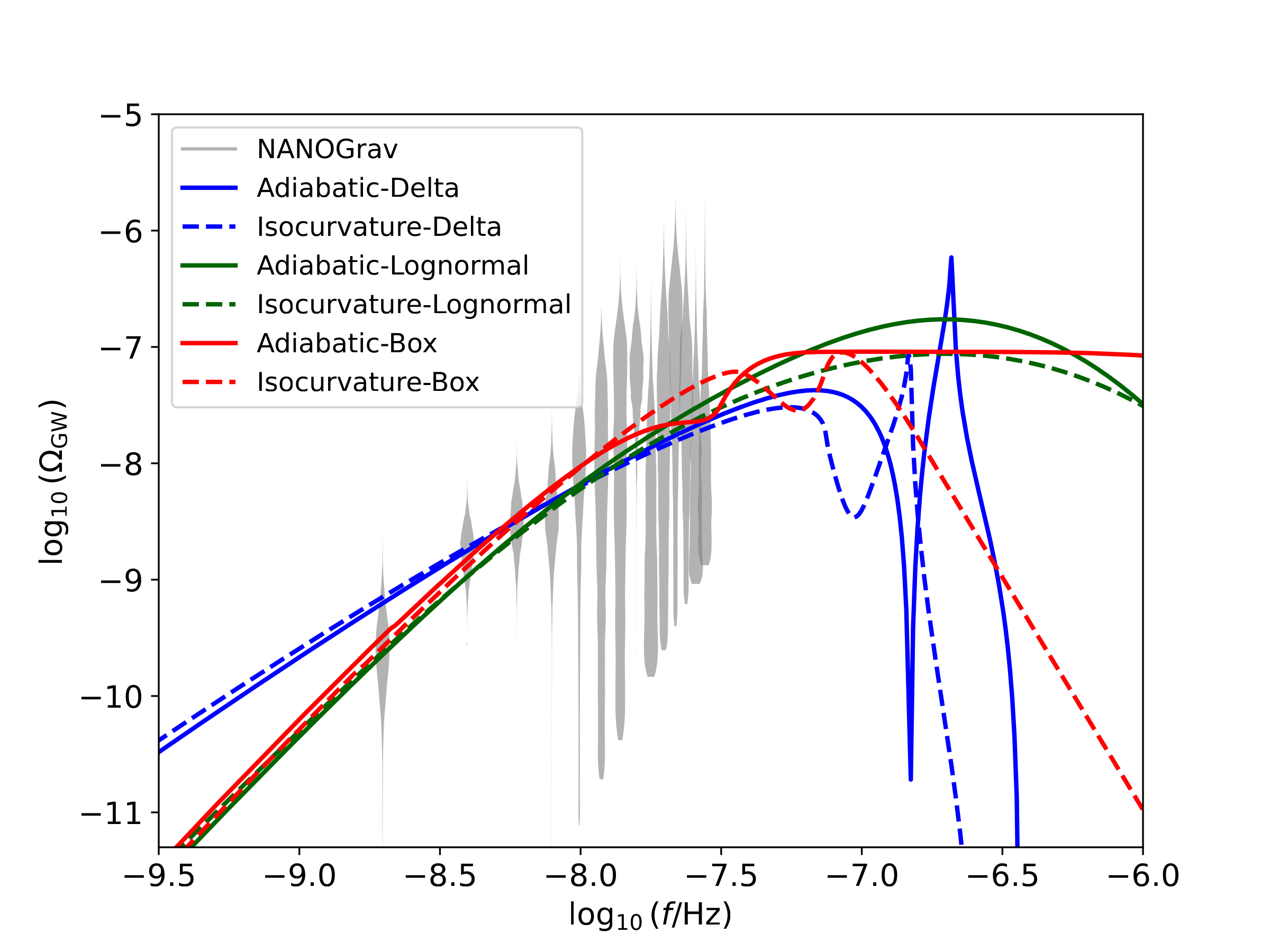}
\caption{{Energy density spectrum of IGWs ($\Omega_{\rm GW}$) as a function of frequency $f$. The gray violin plots represent the free spectral measurements from the NANOGrav 15-year dataset. Colored lines show theoretical predictions for adiabatic (solid) and isocurvature (dashed) scenarios with model parameters fixed to their median posterior values.}}
\label{ogw} 
\end{figure}

To maintain a high level of model independence, we adopt an approach that avoids selecting a specific inflation model capable of generating an enhanced spectrum $\mathcal{P}_{\zeta/s}$. Instead, we focus on three characteristic templates for $\mathcal{P}_{\zeta/s}$ that encompass a range of possibilities typically observed in realistic models. By disregarding the microphysics of inflation, we aim to provide a broader analysis that accommodates various scenarios. The following functional forms of power spectrum are considered:

\begin{itemize}
\item \textsc{Adiabatic/isocurvature delta}: 
\begin{equation}
\label{eq:PRdelta}
\mathcal{P}_{\zeta/s}\left(k\right) = A\,\delta\left(\ln k-\ln k_*\right) \,.
\end{equation}

\item \textsc{Adiabatic/isocurvature lognormal}: 
\begin{equation}
\label{eq:PRgauss}
\mathcal{P}_{\zeta/s}\left(k\right) = \frac{A}{\sqrt{2\pi}\,\Delta}\,\exp\left[-\frac{1}{2}\left(\frac{\ln k -\ln k_*}{\Delta}\right)^2\right] \,.
\end{equation}

\item \textsc{Adiabatic/isocurvature box}: 
\begin{equation}
\label{eq:PRbox}
\mathcal{P}_{\zeta/s}\left(k\right) = A\,\Theta\left(\ln k_{\rm max}- \ln k\right)\Theta\left(\ln k- \ln k_{\rm min}\right) \,.
\end{equation}
\end{itemize}
{An illustration of the energy density spectrum of IGWs as a function of frequency $f$ for different scenarios can be found in \Fig{ogw}.}

\section{Data analyses and results}
\label{Data}


In this study, we analyze the NANOGrav 15-year data set~\cite{NANOGrav:2023hde} to infer the parameters of our model. Our approach particularly leverages the amplitudes of the free spectrum as determined by NANOGrav while accounting for the spatial correlations consistent with the Hellings-Downs~\cite{Hellings:1983fr} curve. The observational sensitivity of a PTA begins at a frequency equivalent to the inverse of the total observation time, $1/T_{\mathrm{obs}}$, where $T_{\mathrm{obs}}=16.03\,\mathrm{yr}$ represents the duration covered by the NANOGrav 15-year data set. We incorporate the posterior distributions of the free spectrum for the $14$ frequency components reported by NANOGrav~\cite{NANOGrav:2023gor} in their analysis of the SGWB signal.

Our analysis begins by examining the posterior time delay data, $d(f)$, as provided by NANOGrav. The relationship between this time delay and the power spectrum, $S(f)$, is
\begin{equation}
S(f) = d(f)^2\, T_{\mathrm{obs}}.
\end{equation}
Utilizing the time delay data allows us to determine the SGWB energy density spectrum by 
\begin{equation}
\hat{\Omega}_{\mathrm{GW}}(f)=\frac{2 \pi^2}{3 H_0^2} f^2 h_c^2(f) = \frac{8\pi^4}{H_0^2} T_{\mathrm{obs}} f^5 d^2(f),
\end{equation}
in which $H_0$ represents the Hubble constant and $h_c(f)$ is the characteristic strain, calculated as
\begin{equation}
h_c^2(f)=12 \pi^2 f^3 S(f).
\end{equation}
At each observed frequency $f_i$, the kernel density estimate, $\mathcal{L}_i$, is derived from the associated $\hat{\Omega}_{\mathrm{GW}}(f_i)$ posteriors. Consequently, the total log-likelihood is the sum across all frequencies of the individual log-likelihoods~\cite{Liu:2023ymk,Wu:2023hsa,Jin:2023wri,Liu:2023pau}:
\begin{equation}
\ln \mathcal{L}(\Lambda) = \sum_{i=1}^{14} \ln \mathcal{L}_i(\Omega_{\mathrm{GW}}(f_i, \Lambda)),
\end{equation}
where $\Lambda$ is a collection of model parameters that will be inferred from the PTA data.

For the parameter space exploration, we utilize the \texttt{dynesty} sampler~\cite{Speagle:2019ivv} incorporated within the \texttt{Bilby} package~\cite{Ashton:2018jfp,Romero-Shaw:2020owr}. The conversion between the wavenumber $k$ and the frequency $f$ is expressed by
\begin{equation}
\label{k-f}
f = \frac{k}{2 \pi} \simeq 1.6\, \mathrm{nHz}
\left(\frac{k}{10^6\,\mathrm{Mpc}^{-1}}\right),
\end{equation}
which provides the necessary link for interpreting spectral features in terms of their spatial scales.
In this work, we consider six models for IGWs: the adiabatic delta model ($M_1$), adiabatic lognormal model ($M_2$), adiabatic box model ($M_3$), isocurvature delta model ($M_4$),  isocurvature lognormal model ($M_5$), and isocurvature box model ($M_6$). In addition, we consider the SMBHB model ($M_0$). The SGWB produced by SMBHBs can be approximated by a power-law spectrum~\cite{Thrane:2013oya}:
\begin{equation}
\Omega_\mathrm{GW}(f) =\frac{2\pi^2 A_\mathrm{SMBHB}^2}{3H_0^2}\left(\frac{f}{f_{\yr}}\right)^{5-\gamma}f_{\yr}^2,
\end{equation}
where $A_{\rm SMBHB}$ is the amplitude of the characteristic strain measured at the reference frequency $f_{\rm yr} \equiv 1\,{\rm yr}^{-1}$, and $\gamma = 13/3$ is the expected power-law spectral index for a population of circular, GW-driven binaries.
The selected priors for the model parameters are comprehensively listed in~\Table{table:priors}. {Moreover, physical constraints are imposed requiring $f_{\rm PBH} \leqslant 1$ to ensure PBH abundance does not exceed dark matter density, as well as that the constraints from CMB \cite{Planck:2018vyg} and BBN \cite{Cooke:2013cba}.}

\begin{figure}[H]
\centering
\includegraphics[width=0.45\textwidth]{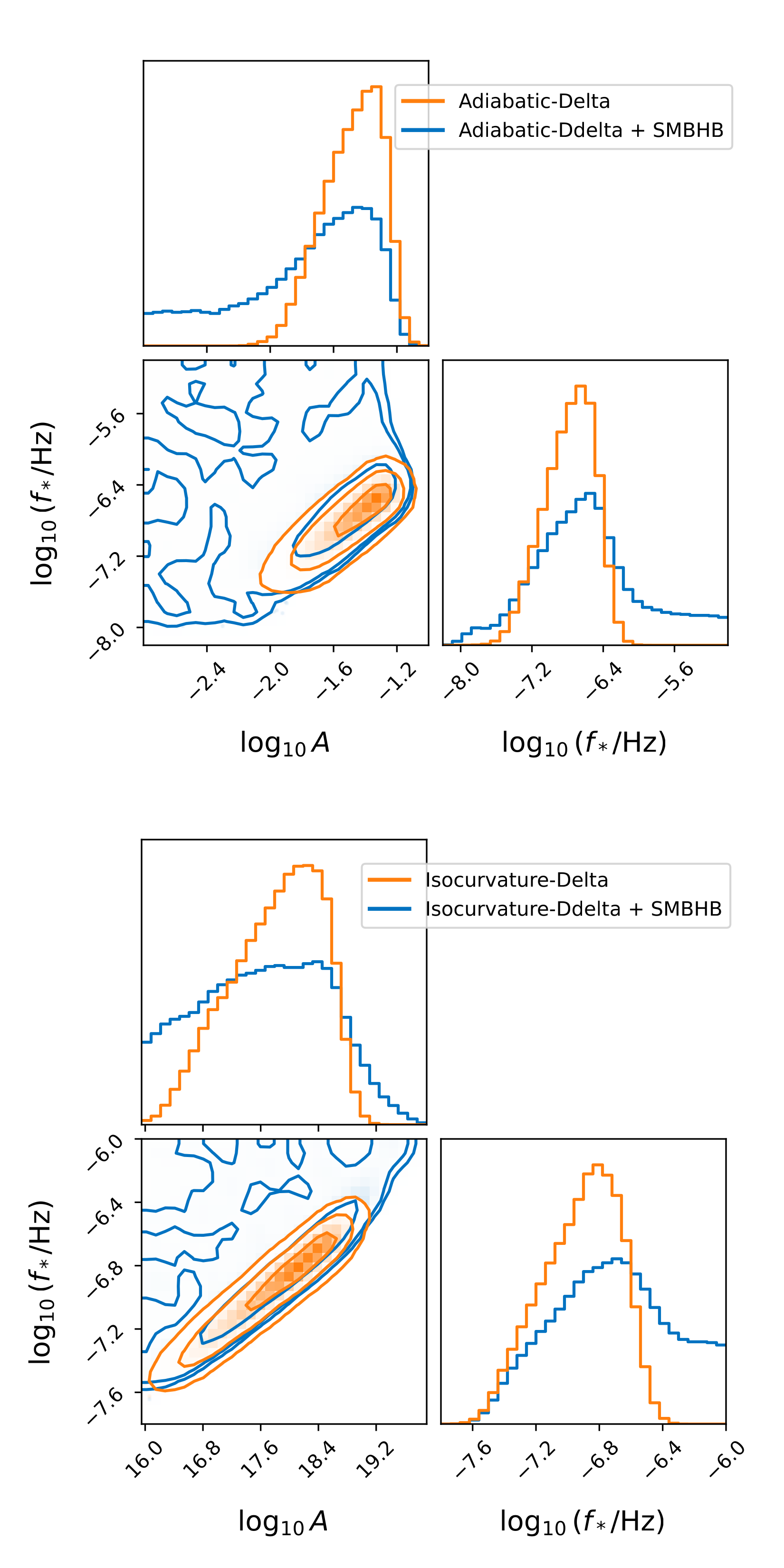}
\caption{Marginalized one- and two-dimensional posterior distributions for model parameters in the \textit{adiabatic delta} (upper panel) and \textit{isocurvature delta} (lower panel) scenarios, analyzed using the NANOGrav 15-year dataset. Blue contours show results for pure IGW models, while orange contours represent the combined IGW+SMBHB analysis. The contours in two-dimensional distributions denote the $68\%$, $95\%$, and $99.7\%$ credible regions ($1\sigma$, $2\sigma$, and $3\sigma$, respectively).}
\label{posts_delta} 
\end{figure}

\begin{figure}[H]
\centering
\includegraphics[width=0.45\textwidth]{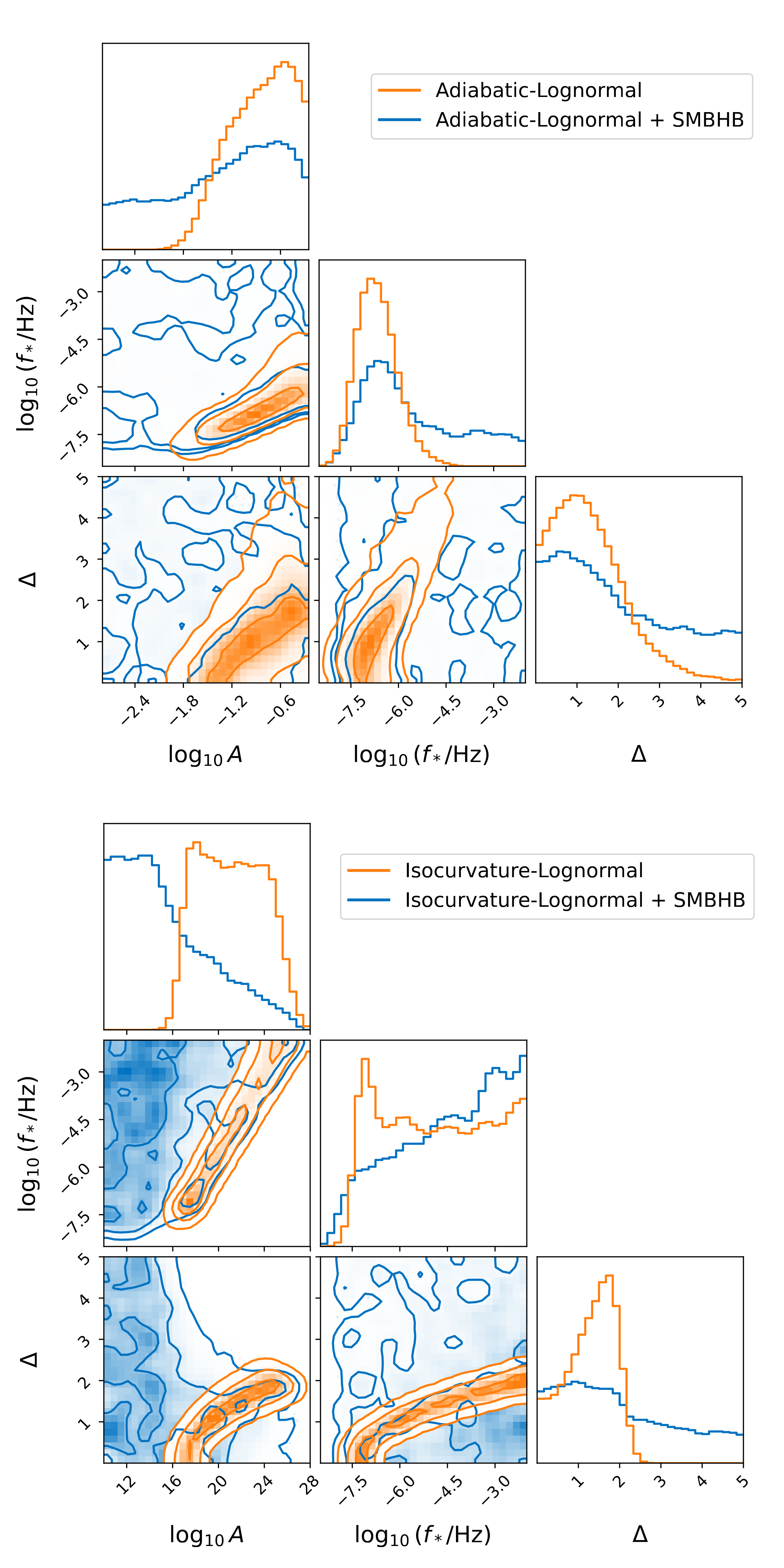}
\caption{Same as \Fig{posts_delta}, but for the \textit{adiabatic lognormal} (upper panel) and \textit{isocurvature lognormal} (lower panel) models.}
\label{posts_log}
\end{figure}

\begin{figure}[H]
\centering
\includegraphics[width=0.45\textwidth]{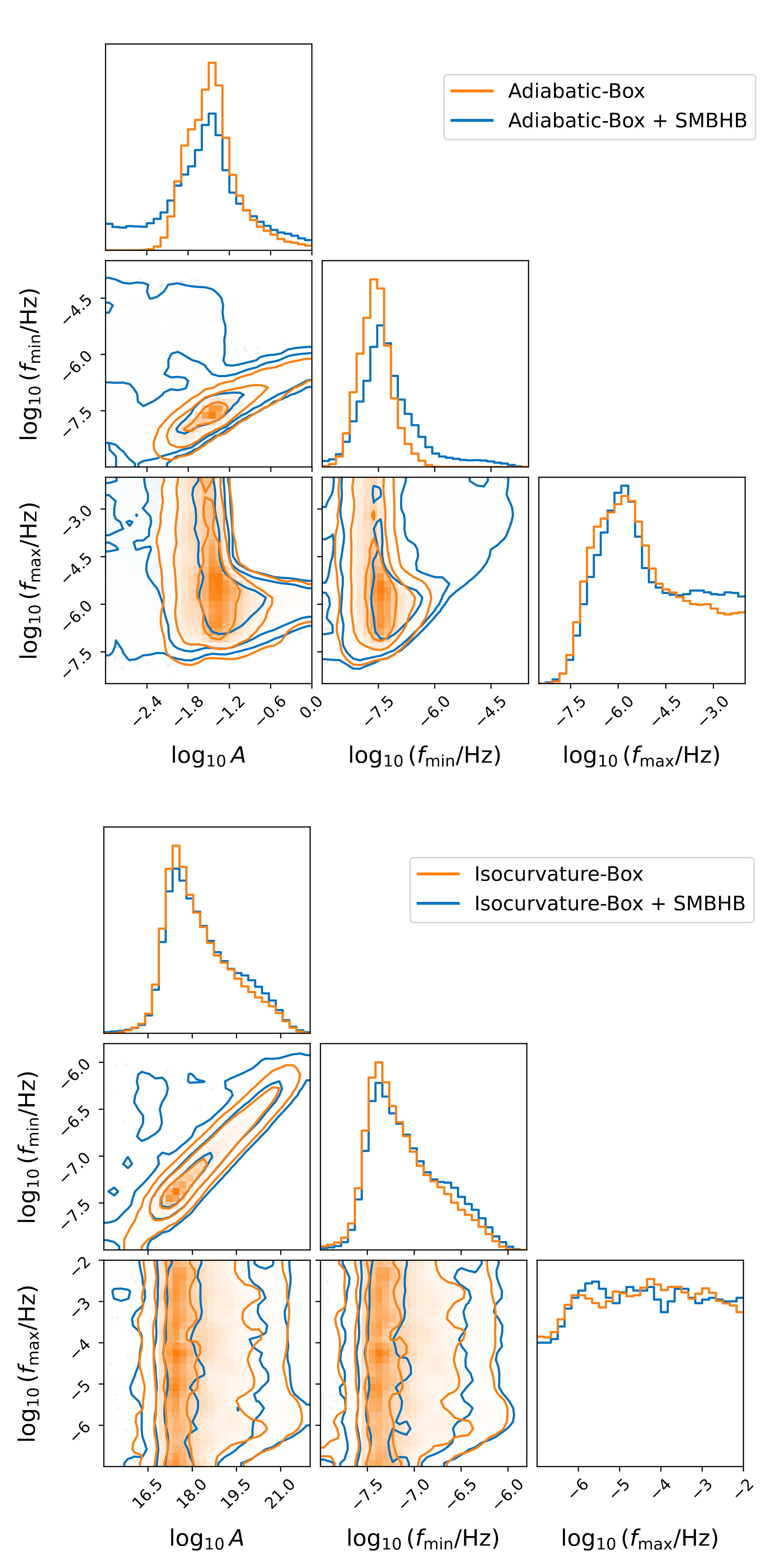}
\caption{Same as \Fig{posts_delta}, but for the \textit{adiabatic box} (upper panel) and \textit{isocurvature box} (lower panel) models.}
\label{posts_box}
\end{figure}

The posterior distributions for the parameters of the models $M_1$ through $M_6$ are illustrated in Figs.~\ref{posts_delta}-\ref{posts_box}. Our analysis indicates that a successful explanation of the NANOGrav signal with IGWs requires a substantial amplitude $A$ for the primordial scalar power spectrum. This inference is supported by computing the lower limits of the $95\%$ Bayesian credible intervals for the amplitude $A$, derived from the one-dimensional marginalized posteriors over the regions of the highest posterior density. For models $M_1$ through $M_6$, we find the following lower limits: $\log_{10} A \gtrsim -1.80$, $\log_{10} A \gtrsim -1.51$, $\log_{10} A \gtrsim -1.93$, $\log_{10} A \gtrsim 16.7$, $\log_{10} A \gtrsim 17.2$, and $\log_{10} A \gtrsim 16.9$, respectively.
Additionally, the amplification in scalar perturbations must be confined to specific scales to ensure that the resulting IGW signal is within the sensitivity range of PTAs. {Additionally, the parameter space is constrained by multiple physical requirements: the avoidance of PBH overproduction, consistency with CMB and BBN bounds.} These constraint imposes specific bounds on the characteristic frequencies $f_\mathrm{min}$, $f_\mathrm{max}$, and $f_*$, which are evident from Figs.~\ref{posts_delta}-\ref{posts_box} and are concisely presented in \Table{table:priors}.

A significant observation is that the posterior distributions for characteristic frequencies, $f_\mathrm{min}$, $f_\mathrm{max}$, and $f_*$, extend to values well above the upper limit of the NANOGrav sensitivity range. This implies that the NANOGrav data align more closely with the low-frequency end of the IGW spectrum. The IGW spectrum continues to rise beyond the NANOGrav frequencies, reaching its peak at $f \gg 1 \mathrm{nHz}$.

\begin{table}[H]
\centering
\caption{Evaluation of model selection scores quantified by the Bayes factor as described in Ref.~\cite{BF}.}
\label{tab:BF}
\begin{tabular}{ccl}
\hline
\hspace{3mm}BF\hspace{3mm} &  Strength of evidence \\
\hline$<1$ &   Negative \\
$1-3$ &  Not worth more than a bare mention \\
$3-20$ &  Positive \\
$20-150$ & Strong \\
$>150$ & Very strong \\
\hline
\end{tabular}
\end{table}

Upon examination, the 2D posterior distributions for the amplitude ($A$) and characteristic frequency ($f_*$) under both delta and lognormal power spectra exhibit a remarkable resemblance. However, it is noted that the distribution corresponding to the lognormal model is marginally wider. This observation aligns with the theoretical understanding that delta and lognormal models are hierarchically related; specifically, the delta power spectrum is a particular instance of the lognormal spectrum as the width parameter $\Delta$ approaches zero. Consequently, the somewhat expanded posterior distribution for $A$ and $f_*$ illustrated in Fig.~\ref{posts_log} mirrors the additional parameter space afforded by the lognormal model, which in turn introduces an incremental degree of parametric freedom. The broader scope of the lognormal model's distribution aptly captures this extra complexity.

To assess which model is most strongly supported by the observational data, we compute the Bayes factor for the competing models. The Bayes factor offers a statistical metric for comparing two mutually exclusive models, $\mathcal{M}_i$ and $\mathcal{M}_j$. It measures the strength of evidence in favor of one model over the other and is defined as
\begin{equation}
\mathrm{BF}_{ij} \equiv \frac{\mathcal{Z}_i}{\mathcal{Z}_j},
\end{equation}
where $\mathcal{Z}_i$ and $\mathcal{Z}_j$ represent the evidence for models $\mathcal{M}_i$ and $\mathcal{M}_j$, respectively. The evidence, denoted by $\mathcal{Z}$, is calculated by integrating the likelihood $\mathcal{L}(\textbf{d} | \Lambda)$ of the data $\textbf{d}$ given the parameters $\Lambda$, weighted by the prior probability distribution $\pi(\Lambda)$ of the parameters, namely
\begin{equation}
\mathcal{Z} \equiv \int \mathrm{d} \Lambda\, \mathcal{L}(\textbf{d} | \Lambda)\, \pi(\Lambda).
\end{equation}
Table~\ref{tab:BF} summarizes the Bayes factor interpretations for model comparison~\cite{BF}.

The calculation of Bayes factors among {seven} competing models yields the following results:
\begin{equation}
\mathrm{BF}_{ij} = \left(
\begin{array}{cccccccc}
& {M_0} & {M_1} & {M_2} & {M_3} & {M_4} & {M_5} & {M_6}\\
{M_0} & {1.00} & {0.12} & {0.16} & {0.09} & {0.17} & {1.23} & {0.08} \\
{M_1} & {8.56} & {1.00} & {1.33} & {0.80} & {1.45} & {10.51} & {0.67} \\
{M_2} & {6.45} & {0.75} & {1.00} & {0.60} & {1.10} & {7.92} & {0.51} \\
{M_3} & {10.70} & {1.25} & {1.66} & {1.00} &  {1.82} & {13.13} & {0.84} \\
{M_4} & {5.89} & {0.69} & {0.91} & {0.55} & {1.00} & {7.23} & {0.46} \\
{M_5} & {0.81} & {0.10} & {0.13} & {0.08} & {0.14} & {1.00} & {0.06} \\
{M_6} & {12.76} & {1.49} & {1.98} & {1.19} & {2.17} & {15.66} & {1.00}
\end{array}
\right).
\end{equation}
{It can be seen that the IGW models show modest preference over the pure SMBHB model, with Bayes factors ranging from $0.81$ to $12.76$. Model $M_6$ exhibits the highest Bayes factor, while model $M_5$ yields the lowest. The Bayes factor comparing these models, $\mathrm{BF}_{65} = 15.66$, indicates positive but not strong evidence favoring $M_6$ over $M_5$ on the Jeffreys scale. The interpretation of the NANOGrav 15-year dataset regarding the common-spectrum process and the associated Hellings-Downs correlation remains subtle, as the collective Bayes factors do not provide decisive evidence for any particular model. The relatively narrow range of Bayes factors among models $M_0$ through $M_6$ precludes a definitive model selection.}

\begin{table}[H]
\centering
\caption{{Bayes factors comparing combined IGW+SMBHB models to their corresponding pure IGW counterparts for scenarios $M_1$ through $M_6$. Each Bayes factor is computed as BF$^{\mathrm{X+SMBHB}}_{\mathrm{X}}$, where X denotes the respective model ($M_1$--$M_6$).}}
\label{tab:BF2}
\begin{tabular}{cc}
\hline
\hspace{3mm}Model\hspace{3mm} & \hspace{3mm}BF$^{\mathrm{X+SMBHB}}_{\mathrm{X}}$\hspace{3mm} \\
\hline
$M_1$& $0.14$ \\
$M_2$& $0.24$\\
$M_3$& $0.15$\\
$M_4$& $0.17$\\
$M_5$& $0.53$\\
$M_6$& $0.54$\\
\hline
\end{tabular}
\end{table}

{We further analyze the statistical evidence for an additional SMBHB component by computing Bayes factors between the combined IGW+SMBHB models and their corresponding pure IGW counterparts. The results are presented in \Table{tab:BF2}. The obtained Bayes factors are consistently less than unity ($\mathrm{BF} < 1$), indicating that the NANOGrav 15-year dataset does not support the presence of an SMBHB contribution beyond the IGW signal. Furthermore, the inclusion of SMBHB components results in increased posterior widths for IGW parameters, an effect expected from the additional degrees of freedom in the combined models.}

\section{\label{Con}Summary and discussion}
In this study, we have investigated the possibility of explaining the SGWB observed by PTAs through IGWs arising from primordial isocurvature and adiabatic fluctuations. By analyzing the latest data from the NANOGrav collaboration, we have computed the Bayes factors comparing various models.
Our findings indicate that the current sensitivity of PTAs is insufficient to distinguish between isocurvature and adiabatic fluctuations. The analysis of the NANOGrav data, while providing valuable insights into the nature of the SGWB, does not allow us to conclusively determine the origin of the observed PTA signal. This result highlights the challenges involved in unraveling the primordial origins of the SGWB and the need for further advancements in sensitivity and data analysis techniques.

While our current analysis suggests that PTAs cannot yet distinguish between isocurvature and adiabatic fluctuations, we anticipate that future advancements in observational capabilities and data analysis techniques will gradually improve our ability to probe the primordial origins of the SGWB. The continued progress in constructing next-generation radio telescopes, such as the Square Kilometre Array~\cite{Lazio:2013mea}, promise to provide unprecedented sensitivity and frequency coverage, enabling more precise measurements of the SGWB and its properties. {It is crucial to recognize that our analysis is based on the assumption of either purely adiabatic or purely isocurvature perturbations. However, the total fluctuations in reality may comprise a combination of both types of perturbations. While incorporating a ratio between the contributions of adiabatic and isocurvature perturbations to the IGW energy-density spectra is beyond the scope of the current study, it presents a promising direction for future research. We leave the exploration of mixed scenarios for subsequent studies, as they may offer valuable insights into the nature of primordial perturbations.}

Furthermore, ongoing efforts to refine theoretical models of the early Universe and the generation of primordial fluctuations will contribute to our understanding of the observed PTA signal. Refining our understanding of the nonlinear physics involved in the evolution of small-scale fluctuations will be crucial for accurately modeling the IGWs and distinguishing them from other sources contributing to the SGWB.

In conclusion, our study emphasizes the importance of the SGWB as a powerful probe of the early Universe and highlights the potential of PTAs in unraveling the primordial origins of GWs. While the current sensitivity of PTAs is insufficient to distinguish between isocurvature and adiabatic fluctuations, future advancements in observational capabilities and theoretical modeling, combined with multi-PTA and multi-messenger approaches, hold the promise of shedding light on the nature and origin of the observed PTA signal.

\Acknowledgements{Zu-Cheng Chen is supported by the National Natural Science Foundation of China under Grant No.~12405056 and the innovative research group of Hunan Province under Grant No.~2024JJ1006. Lang Liu is supported by the National Natural Science Foundation of China Grant under Grant No.~12433001. }

\InterestConflict{The authors declare that they have no conflict of interest.}
\bibliographystyle{JHEP}
\bibliography{Ref}
\end{multicols}
\end{document}